\documentclass[aps,prl,preprint,groupedaddress]{revtex4}
\usepackage{graphics,epsfig}
\usepackage{axodraw}

\begin{document}

\preprint{SHEP-07-04}

%Title of paper
\title{\Large\bf Explicit CP Violation in the MSSM \\
Through Higgs $\rightarrow \gamma\gamma$}

\author{S. Moretti}
\email{stefano@hep.phys.soton.ac.uk}
\author{S. Munir}
\email[]{shobig@hep.phys.soton.ac.uk}
\author{P. Poulose}
\email[]{poulose@hep.phys.soton.ac.uk}
\altaffiliation[Also at ]{IIT Guwahati, INDIA; poulose@iitg.ernet.in}
%\thanks{On leave from IIT Guwahati, INDIA - 781 039}
\affiliation{School of Physics \& Astronomy,\\ 
University of Southampton,  Highfield, Southampton SO17 1BJ, UK}
%\homepage[]{Your web page}
%\altaffiliation{}

\date{\today}

\begin{abstract}
The MSSM with explicit CP violation is studied through the di-photon decay
channel of the lightest neutral Higgs boson. Through the leading one-loop order
$H_1\rightarrow \gamma\gamma$ is affected by a large number of 
Higgs-sparticle
couplings, which could be complex. Our preliminary scan over the
Supersymmetric parameter space shows that more than 50\% average deviations 
are possible, in either direction, in the corresponding branching ratio, with 
respect to the case of the CP-conserving MSSM. In particular, our analysis shows that 
in the presence of a light stop (with mass $\sim 200$ GeV) a CP-violating phase 
$\phi_\mu \sim 90^\circ$ can render the $H_1\rightarrow \gamma\gamma$ branching 
ratio more than 10 times larger, for suitable combinations of the other MSSM
parameters.
\end{abstract}

% insert suggested PACS numbers in braces on next line
%\pacs{}
% insert suggested keywords - APS authors don't need to do this
%\keywords{}

%\maketitle must follow title, authors, abstract, \pacs, and \keywords
\maketitle

\newcommand{\ra}{\rightarrow}
\newcommand{\lra}{\longrightarrow}
\newcommand{\ee}{e^+e^-}
\newcommand{\gam}{\gamma \gamma}
\newcommand{\tb}{\tan \beta}
\newcommand{\s}{\smallskip}
\newcommand{\nn}{\noindent}
\newcommand{\non}{\nonumber}
\newcommand{\beq}{\begin{eqnarray}}
\newcommand{\eeq}{\end{eqnarray}}

\providecommand{\UI}{}
\renewcommand{\UI}{\ensuremath{\mathrm{U}(1)}}
\providecommand{\smgroup}{}
\renewcommand{\smgroup}{\ensuremath{\mathrm{SU}(3) \otimes \mathrm{SU}(2)_L \otimes \mathrm{U}(1)_Y } }
\providecommand{\Tr}{}
\renewcommand{\Tr}{\ensuremath{\mathrm{Tr}}}
\providecommand{\identity}{}
\renewcommand{\identity}{\ensuremath{\mathrm{I}}}
\providecommand{\Kahler}{}
\renewcommand{\Kahler}{K\"ahler}
\providecommand{\Fbar}{}
\renewcommand{\Fbar}{\overline{F}}

\noindent
The mechanism of Electro-Weak Symmetry Breaking (EWSB) is elusive even
after the very successful LEP era, although precision measurements hint
at a light Higgs particle. It is expected that the soon to be operational Large 
Hadron Collider (LHC) will
be able to make definite statements about the Higgs mechanism.
At the same time there are various reasons to think that the Standard Model 
(SM) is only an effective theory valid up to TeV range, and some richer
structure is needed to explain particle dynamics (much) beyond such energy scale. 
Supersymmetry (SUSY), being the most favoured of all the new physics scenarios
proposed so far, is going to be searched for in all possible ways at the LHC.
Within the Minimal Supersymmetric Standard Model (MSSM) the scalar potential 
conserves CP at tree level \cite{Higgs-hunter}. The reason is that
SUSY imposes an additional (holomorphic) 
symmetry on the Higgs sector of a general two-Higgs doublet model, that 
entails flavour conservation in tree-level neutral currents and absence of 
CP-violating scalar-pseudoscalar mixings in the Born approximation. Beyond the 
latter, recent studies have shown that CP invariance of the Higgs 
potential may in principle be broken by radiative corrections \cite{Maekawa}, 
as the Vacuum Expectation Values (VEVs) of the two Higgs doublets can develop a 
relative phase \cite{Pilaftsis-PLB}. This type of CP violation is generally 
referred to as spontaneous CP violation and it requires a light Higgs 
state as a result of the Georgi-Pais theorem 
\cite{Georgi-Pais}, but the possibility of the latter has now been 
essentially ruled out by experiment \cite{Pamoral}.

CP violation can also be explicitly induced in the MSSM, in much the same way 
as it is done in the SM, by complex Yukawa couplings of the Higgs bosons to 
(s)quarks. There are several new parameters 
in the Supersymmetric theory, that are absent in the SM, which could well be 
complex and thus possess CP-violating phases. Such parameters include: (i) the 
higgsino mass term $\mu$, (ii) the soft SUSY-breaking gaugino masses 
$M_a~(a = 1,2,3)$, (iii) the soft bilinear term $B\mu$ 
and (iv) the soft trilinear 
Yukawa couplings $A_f$ of the Higgs particles to scalar fermions of flavour
$f$. 

Each of these parameters can have independent phases.  After applying universality 
conditions at a unification scale $M_X$ the gaugino masses have a common phase and the 
trilinear couplings are  all equal with another common phase. As argued by
\cite{dchang}, one may deviate from exact universality and consider $A_f$ 
to be diagonal in flavour space with vanishing first and second generation
couplings to avoid problems with the electron, muon and neutron Electric
Dipole Moments (EDMs).
This leaves four independent phases, those of  $\mu,B\mu,M_a$ and $A_f$. 
However, the two $U(1)$ symmetries of the conformal-invariant part of the
MSSM may be employed to re-phase one of the Higgs doublet fields and the
gaugino fields such that $M_a~\rm{and}~\textit{B}\mu$ are real 
\cite{PlWagner,Dugan}. We will work within this setup with two independent
physical phases, which we take to be $arg(\mu)=\phi_\mu$ and $arg(A_f)=\phi_{A_f}$. As intimated, 
the CP-violating phases associated with the sfermions of the first and, to a 
lesser extent, second generations are severely constrained by bounds on the 
EDMs of the electron, neutron and muon. However, 
there have been several suggestions \cite{EDM1}--\cite{EDM3} to evade these 
constraints without suppressing the CP-violating phases. One possibility is to 
arrange for partial cancellations among various contributions to the 
EDMs \cite{EDM3}.
Another option is to make the first two generations 
of scalar fermions rather heavy, of order a few TeV, so that the one-loop 
EDM constraints are automatically evaded. As a matter of fact, one can consider
so-called effective SUSY models \cite{EDM2} where decoupling of the first 
and second generation sfermions are invoked to solve the SUSY Flavour Changing 
Neutral Current (FCNC) and CP 
problems without spoiling the naturalness condition. We adopted the latter 
version of a CP-violating MSSM for our analysis, along with $A_f=0$ for 
the first two generation sfermions. 

\begin{figure}
\begin{center}
\setlength{\unitlength}{1.0pt}
\begin{picture}(400,100)(-20,0)
\Photon(50,20)(80,20){-3}{4}
\Photon(50,80)(80,80){3}{4}
\ArrowLine(50,20)(50,80)
\ArrowLine(50,80)(20,50)
\ArrowLine(20,50)(50,20)
\DashLine(-10,50)(20,50){5}
\put(-25,46){$H_i$}
\put(60,46){$f,\tilde \chi^\pm$}
\put(85,18){$\gamma$}
\put(85,78){$\gamma$}
\Photon(190,20)(220,20){-3}{4}
\Photon(190,80)(220,80){3}{4}
\DashLine(190,80)(190,20){5}
\DashLine(190,20)(160,50){5}
\DashLine(160,50)(190,80){5}
\DashLine(130,50)(160,50){5}
\put(115,46){$H_i$}
\put(195,46){$W, H^\pm, \widetilde{f}$}
\put(225,18){$\gamma$}
\put(225,78){$\gamma$}
\DashLine(280,50)(310,50){5}
\DashCArc(325,50)(15,0,180){4}
\DashCArc(325,50)(15,180,360){4}
\Photon(340,50)(375,80){3}{5}
\Photon(340,50)(375,20){3}{5}
\put(265,46){$H_i$}
\put(305,70){$W, H^\pm, \widetilde{f}$}
\put(380,18){$\gamma$}
\put(380,78){$\gamma$}
\end{picture}
\end{center}
\caption{Diagrams for Higgs decay into $\gamma\gamma$ pairs in the CP-violating MSSM:
$f\equiv t,\: b;\;\;\tilde f\equiv \tilde t_{1,2}, \tilde b_{1,2},
\tilde \tau_{1,2}$.}
\label{fig:HiggsPho}
\end{figure}
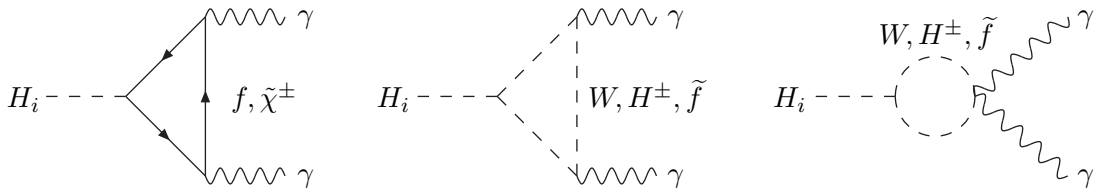

The CP-violating phases $\phi_\mu$ and $\phi_{A_f}$ could
in principle be measured directly in the production cross sections and
decay widths of (s)particles in high energy colliders \cite{PlWagner},
\cite{ChoiSik} - \cite{CarEllis1a} or indirectly via their radiative effects
on the Higgs sector \cite{PlWagner, ChoiKao}.
In this letter we will look at $H_1\rightarrow \gamma\gamma$ which involve
the (leading) direct effects of CP violation through couplings of $H_1$ to
sparticles in the loops (see Fig.~\ref{fig:HiggsPho}) as well
as the (subleading) indirect effect through the scalar-pseudoscalar mixing yielding  
a CP-mixed state, $H_1$. The origin of this CP-mixing is the following.
In the Higgs sector,
the CP-violating phases mentioned above introduce non-vanishing
off-diagonal mixing terms in the neutral Higgs mass matrix, which in the
weak basis $(\phi_1,\phi_2,a)$, where $\phi_{1,2}$ are the  CP-even states 
and $a$ is the CP-odd state, may schematically be written as 
\cite{PlWagner,ChoiKao,CarEllis1,heinmeyer}
\beq
\mathcal{M}_H^2 = \left(%
\begin{array}{cc}
  \mathcal{M}_S^2 & \mathcal{M}_{SP}^2 \\
  \mathcal{M}_{PS}^2 & \mathcal{M}_P^2 \\
\end{array} \right).
\eeq
Here, ${\cal M}_S^2$ is a $2\times 2$ matrix describing the transition between 
the CP-even states, ${\cal M}_P^2$ gives the mass of the CP-odd
state whilst ${\cal M}_{PS}^2=({\cal M}_{SP}^2)^T$ (a $1\times 2$ matrix) 
describes the mixing between the CP-even and CP-odd states. The mixing matrix 
elements are typically proportional to 
\beq
{\cal M}_{SP}^2\propto \mathcal{I}m (\mu A_f)
\eeq
and are dominated by loops involving the top squarks 
and could be of order $M_Z^2$. 
As a result, the neutral Higgs bosons of the MSSM no longer carry any definite 
CP-parities. Rotation from the EW states to the mass eigenvalues, 
\beq
(\phi_1,\phi_2,a)^T=O\;(H_1,H_2,H_3)^T, \non
\eeq
is now carried out by a $3\times 3$
real orthogonal matrix $O$, such that
\beq 
O^T\mathcal{M}_H^2O = {\rm{diag}}(M^2_{H_1},M^2_{H_2},M^2_{H_3}) 
\eeq
with $M_{H_1}\leq M_{H_2}\leq M_{H_3}$. 
As a consequence, it is now appropriate to parameterise the Higgs sector of the CP-violating MSSM in terms of
the mass of the charged Higgs boson, $M_{H^\pm}$, as the latter remains basically unaffected. (For a detailed 
formulation of the MSSM Higgs sector with explicit CP violation, see Refs. \cite{PlWagner,CarEllis1}.) 

In order to study the effects of the CP-violating phases we focus here on the 
di-photon decay mode of the lightest neutral Higgs boson, $H_1$. The reason is twofold.
Firstly, the di-photon decay mode is 
the most promising channel for the discovery of a light neutral Higgs state --
of mass between, say, $80-130$ GeV, at the LHC \cite{ATLASTDR2,CMS}. Secondly,
the dominant CP-violating terms dependent on $\mu$ and $A_f$ (hereafter, $f=b,t,\tau$)
enter the perturbative calculation of the di-photon decay width 
with a coupling strength that is of the same order as that of the
CP-conserving ones (of ${\cal O}(\alpha^3)$). Furthermore, on the technical side, 
thanks to the narrow width of such a light Higgs state (of 10 MeV at the most), the entire 
$gg/qq\ra H_1\ra\gamma\gamma$ process can be factorised into three parts: 
the production process, the Higgs propagator and the decay channel. 
Effects of CP violation can show up in
this process through the aforementioned couplings in the production, through a possible
mixing of Higgs states at one-loop and above in the propagator and through
the same couplings in the decay. 
%\cite{noteEllisPLee}. 
CP violation entering the production of a Higgs state in gluon-gluon 
fusion process at hadron colliders was studied first by \cite{dedes}, 
choosing a parameter space region which is not sensitive to the CP-mixing of 
the Higgs states, and later by \cite{ChoiSik2,ChoiKaoJae}, including the presence 
of CP-mixing of the Higgs states.  Effects of CP-mixing in the propagator are 
discussed separately but in great detail in \cite{EllisPLee}. 
%Here, we will concentrate on the decay of a physical Higgs particle, $H_1$, which is now a CP-mixed state, into two photons. 
A thorough study of the other MSSM Higgs decay channels in presence
of CP-violation can be found in 
\cite{ChoiSik}-\cite{CarEllis1},\cite{dedes}-\cite{dilip}.
%%{\bf again, dig out the Choi and/or Hagiwara paper where this was done, plus Pilaftsis' one}.
We postpone the full analyses of $gg/qq\ra H_1\ra\gamma\gamma$ including also 
CP-violating effects in production and propagation to a later work. 

In the MSSM a Higgs state decays into two photons through loops of fermions, 
sfermions, EW gauge bosons, charged Higgses as well as charginos  
(see Fig. \ref{fig:HiggsPho}). 
(Expressions for the amplitudes of $H_1\ra \gamma\gamma$ along with 
relevant couplings are available in \cite{CPSuperH} and references therein.)
This decay mode of $H_1$ (also $H_2$ and $H_3$) along with its production 
through gluon-gluon fusion 
is discussed by Ref.~\cite{ChoiKaoJae}. However, the analysis therein was 
limited to parameter space regions where CP-violating effects are only due to 
the changed SM particle (especially $W^\pm$) couplings to the $H_1$. 
Effects of sparticles (made suitably heavy in Ref.~\cite{ChoiKaoJae}) 
in the loops were negligible. 
We examine here the complementary region of MSSM parameter space (albeit limitedly
to the Higgs decay), wherein sparticle masses are taken light, so that they
contribute substantially in the loops. In addition,
we will show that, in the presence of non-trivial CP-violating phases, there
are regions of MSSM parameter space where the couplings of the 
decaying Higgs bosons to all sparticles in the loops are strongly modified 
with respect to the CP-conserving MSSM, thereby inducing dramatic changes
on the $H_1\to\gamma\gamma$ width and Branching Ratio (BR). 

To prove this, we have used the publicly available {\tt FORTRAN} code {\tt CPSuperH} \cite{CPSuperH} version 2 
for our analysis. {\tt CPSuperH} calculates the mass spectrum and decay widths of the neutral and charged Higgs bosons 
in the most general MSSM including explicitly CP-violating phases. In addition, it computes all the couplings of the 
neutral Higgs bosons $H_{1,2,3}$ and the charged Higgs boson $H^\pm$ to ordinary and Supersymmetric matter. The program 
is based on the results obtained in Refs.~\cite{ChoiSik}--\cite{ChoiDrees} and the most recent renormalisation 
group improved effective-potential approach, which includes dominant higher-order logarithmic and threshold corrections, 
$b$-quark Yukawa-coupling resummation effects and Higgs boson pole-mass shifts \cite{CarEllis1,CarEllis2}. 
%CPSuperH efficiently computes the neutral and charged Higgs boson couplings and masses with equally high levels of precision in the CP-conserving case also \cite{HPheno}. 

The free non-SM parameters of the model now include:
$|\mu|$, phase of $\mu~(\phi_{\mu})$, charged Higgs mass ($M_{H^\pm}$), 
soft gaugino masses ($M_a$), 
soft sfermion masses of the third generation ($M_{(\tilde Q_3,\tilde U_3, \tilde D_3, \tilde L_3, \tilde E_3)}$), 
(unified) soft trilinear couplings of the third generation 
($|A_f|$), phase of the trilinear coupling 
($\phi_{A_f}$).  In our scan we have chosen the following very extensive parameter ranges:

\vskip 5mm
\noindent
$$\tan\beta: 1 - 60,\;
|\mu|: 100 - 2000\; {\rm GeV},\;
\phi_{\mu}: 0^\circ - 180^\circ,\; 
M_{H^\pm}: 100 - 400\; {\rm GeV},\;$$
$$
M_2: 100 - 500\; {\rm GeV},\;
M_{(\tilde Q_3, \tilde U_3, \tilde D_3, \tilde L_3, \tilde E_3)}: 100 - 2000\;{\rm GeV},\;
|A_f|: 100 - 2000\; {\rm GeV}.$$
\vskip 5mm

We aimed at searching regions in the Supersymmetric parameter space where the variation 
in the BR$(H_1\to\gamma\gamma)$ due to the CP-violating phases is maximised 
compared to the CP-conserving case.  As stated above, the CP-violating effects 
are proportional to ${\cal I}m (\mu A_f)$, and so we opted to fix 
$\phi_{A_f}$ to 0$^\circ$ and varied only $\phi_\mu$. Besides, $M_1~{\rm{and}}~M_3$ were 
kept fixed as their variation is of no significance here. 
Finally, no unification of the soft sfermion masses was assumed, though they were taken in the same range. 
For this analysis, threshold corrections induced by the exchange of gluinos and charginos in the Higgs-quark-antiquark 
vertices \cite{Hemph,Coarasa} were not included. 

We scanned the above parameter space for 100,000 randomly selected points 
within the ranges specified above and for each of these we have taken values 
of $\phi_\mu$ increasing from $0^\circ$ to $180^\circ$ in steps of $20^\circ$.
Notice that $\phi_\mu=0^\circ(180)^\circ$ corresponds to the CP-conserving 
MSSM point with $\mu=+|\mu|(-|\mu|)$, while any other non-trivial $\phi_\mu$ 
shows the effect of CP violation.  The following experimental constraints from 
LEP2 and Tevatron \cite{PDG,Grivaz} were imposed during the scan:   
\beq
m_{\chi_1^\pm} &\geq& 104~\textrm{GeV}~\textrm{(LEP2)}, \non\\
m_{\tilde f} &\gtrsim& 100~\textrm{GeV for}~\tilde f = \tilde l, \tilde \nu, \tilde t_1~\textrm{(LEP2)}, \non\\
m_{\tilde b} &\gtrsim& 300~\textrm{GeV (Tevatron)}. \non
\eeq
Parameter space points violating these constraints were discarded. 
We only considered points with lightest Higgs mass ($M_{H_1}$) between 90 and 
130 GeV, the range in which the $H_1 \rightarrow \gamma\gamma$ decay is relevant.

\begin{figure}
\includegraphics[width=8cm]{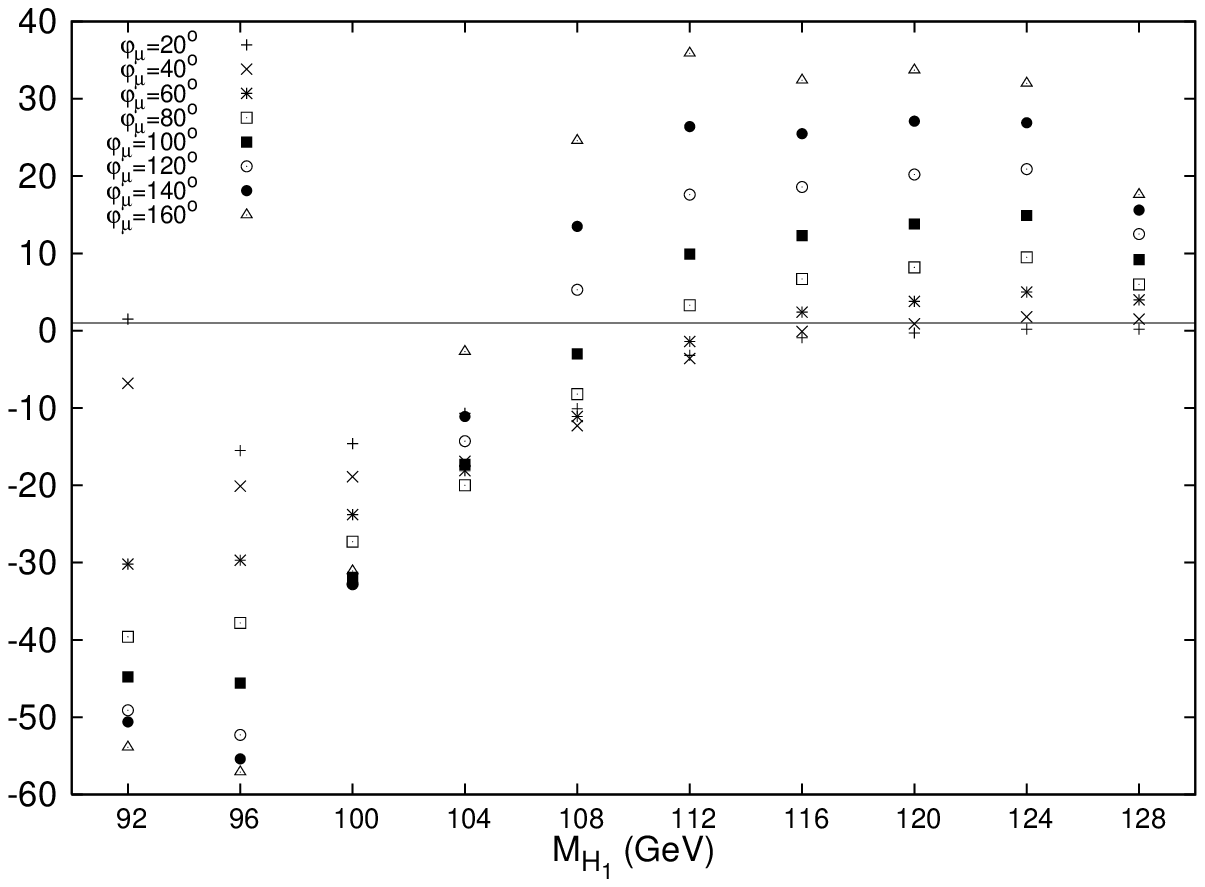}
\includegraphics[width=8cm]{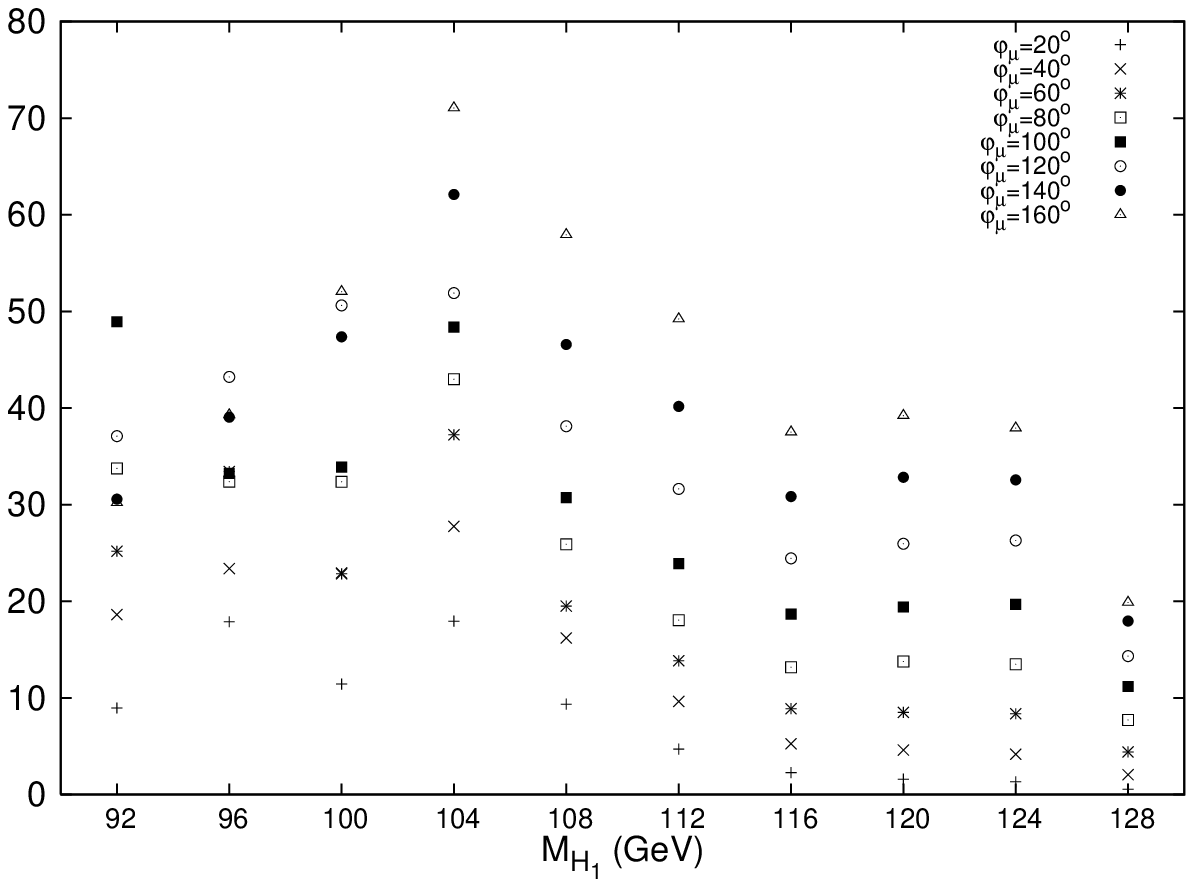}
\caption{
Scan result showing binned  average values of percentage differences 
in the BR($H_1\ra \gamma\gamma$) between the CP-violating and  CP-conserving case (left) as
well as the absolute value of it (right) -- see Eq.~(\ref{eq:BRRatio}) -- for various 
choices of $\phi_\mu$. 
}
\label{fig:BRRatio}
\end{figure}

For each point in the scans that survives the various constraints discussed 
above  we asked {\tt CPSuperH} to print out the mass and $\gamma\gamma$  
BR of the lightest Higgs $H_1$.
In order to have an idea of the overall trend followed by the BR for 
different phases, we first looked at the average behaviour at specific 
$M_{H_1}$ values. To do this we divided the mass range into bins of size 4
GeV.  To find the average sensitivity within each mass bin, we defined the 
percentage deviation
\beq
R^i_{\phi_\mu} = \frac{\sum_n 
\left({\rm{BR}}^{(i,n)}_{\phi} - {\rm{BR}}^{(i,n)}_{0} \right)}
{\sum_n {\rm{BR}}^{(i,n)}_{0}}\times 100,
\label{eq:BRRatio}
\eeq
where the summation is over the number of random points ($n$) 
within a particular bin $i$. We denote the BR 
in the CP-conserving case by ${\rm{BR}}_0$ (specifically, the latter corresponds to the case
$\phi_\mu=0^\circ$, however, without any loss or gain of information, we could alternatively have used the 
limit $\phi_\mu=180^\circ$)
 and that with a non-vanishing $\phi_\mu$ (different from $180^\circ$)
by ${\rm{BR}}_\phi$. This average percentage deviation is plotted in Fig. 
\ref{fig:BRRatio} (left) for the different values of $\phi_\mu$ taken in each 
bin.  
There is an enhancement in the BR of about 20\% for $M_{H_1}$ larger than
110 GeV for moderate values of $\phi_\mu \sim 100^\circ$, while there is a 
suppression of about $40 - 50\%$ for $M_{H_1}$ around 90 -- 98 GeV.
This change-over from enhancement to suppression for lower $M_{H_1}$
values shows the diminishing role of sparticles in the loop as the 
mass difference between $2m_{\tilde f}$ and $M_{H_1}$ increases and the effect of a
non-zero $\phi_\mu$ is effectively more and more through a changed $H_1WW$ coupling. Such
suppression is in agreement with the results of \cite{ChoiKaoJae}. 
In the mass region of 100 -- 110 GeV the effect is apparently very small.
However, this is an artifact of the binned averaging, where points with enhanced 
and suppressed BRs falling in the same mass bin cancel each other. This 
cancellation is nullified by taking the absolute value of the difference in the numerator of Eq.~(\ref{eq:BRRatio}), i.e., 
before averaging. The result is plotted in Fig.~\ref{fig:BRRatio} (right).
More than 50\% deviation is seen for $\phi_\mu=100^\circ$ for $M_{H_1}$ 
around 104 GeV.
Now, it should be noted that these figures represent only the average 
behaviour. It is therefore possible to find regions of parameter space 
where the differences are larger (or smaller, for that matter). We did 
indeed find points with difference in the BR larger than 50 times in our scan
in either direction. 

A subtlety should be noted
in this context though, concerning the derived MSSM masses that also depend on $\phi_\mu$ and 
enter the decay $H_1\to\gamma\gamma$ ($M_{H_1}$, $m_{\tilde b,\tilde t,\tilde \tau}$, $M_{\chi^\pm}$).
In fact, all the latter change when going from the CP-conserving case to the CP-violating one. The most
crucial one in this respect is $M_{H_1}$. However, we have verified that for the same parameter point (apart from
a different $\phi_\mu$)  the latter always changes less than 2 GeV between the two MSSM configurations. Hence,
our 4 GeV wide bins do capture percentage corrections to the BR consistently between the two MSSMs as a function
of the lightest Higgs boson mass. (Rare borderline cases are also correctly assigned to the right bin.) 
Besides, 2 GeV is roughly the di-photon mass resolution in ATLAS 
\cite{ATLASTDR2} (while in CMS it is somewhat better \cite{CMS}).
In short, we imagine an experimental situation in
which a Higgs resonance is extracted in $\gamma\gamma X$ samples with the above mass resolution at a time
when the other SUSY masses and mixing (including $\tan\beta$) entering the loops of the $H_1\to\gamma\gamma$
mode have already been measured in real sparticle production with a resolution that does not allow one
to distinguish between a CP-conserving and a CP-violating MSSM scenario.
Under these conditions, for large enough differences of BRs,
a simple measurement of the normalisation of the $\gamma\gamma$ resonance (after
background subtraction) may suffice to distinguish between the two envisaged 
CP scenarios. 

\begin{figure}
\includegraphics[width=8cm]{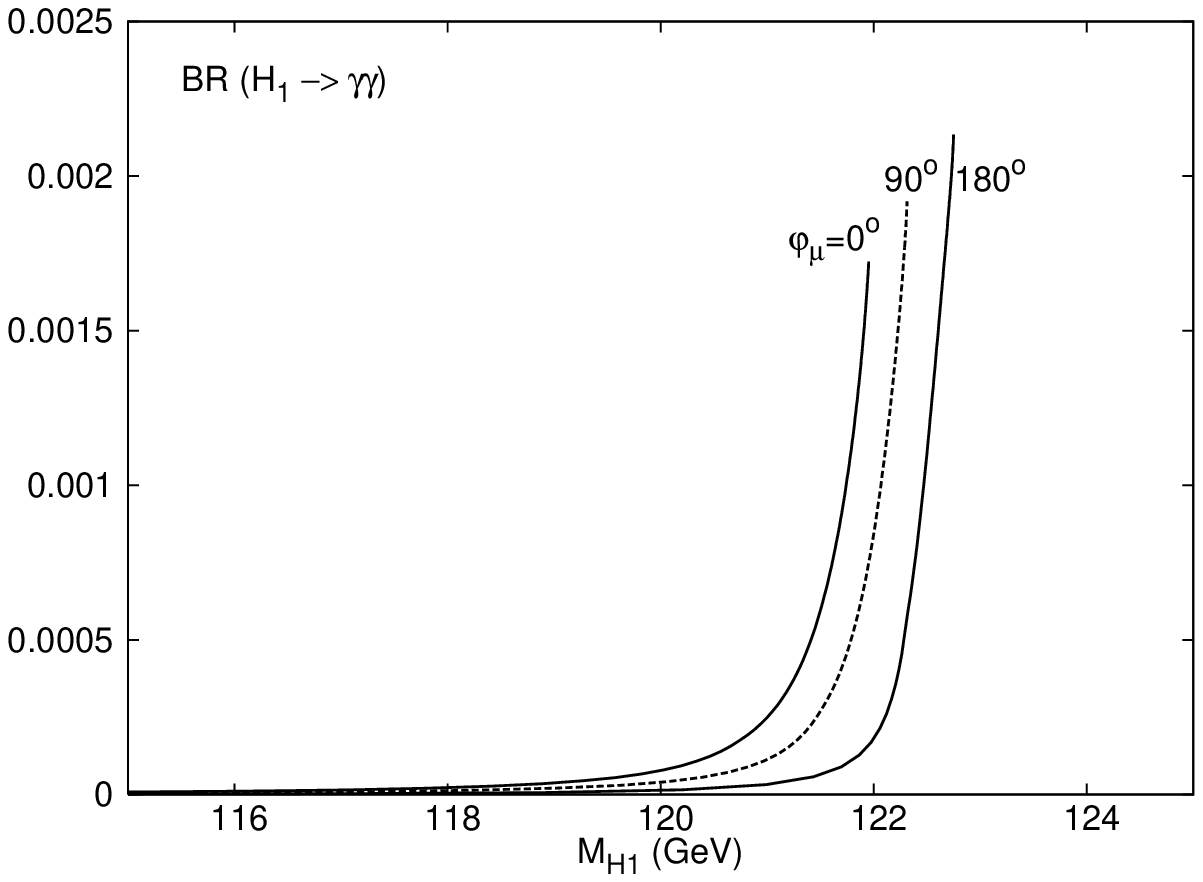}
\includegraphics[width=8cm]{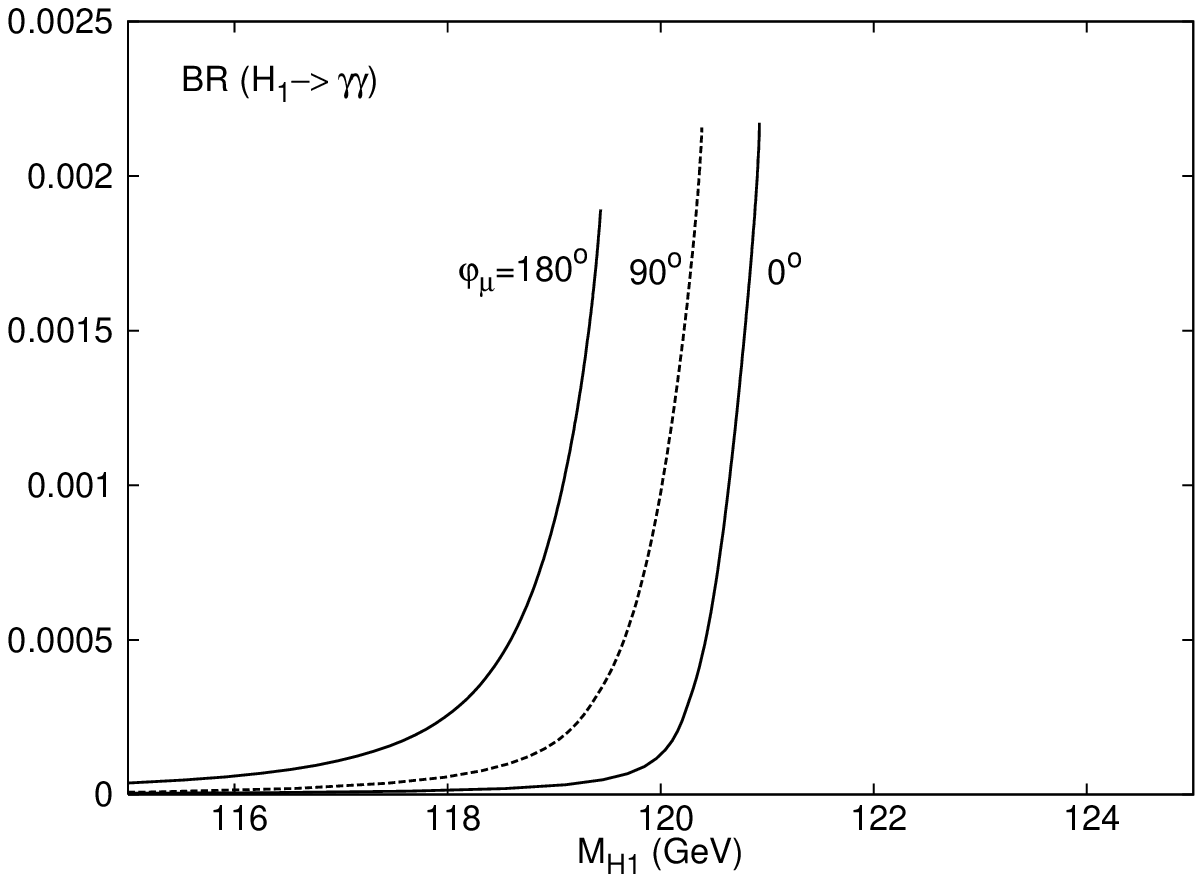}
\caption{BR($H_1\ra \gamma\gamma$) plotted against $M_{H_1}$.
Parameters used are: 
$\tan\beta=20,\;\; M_1=100\;{\rm GeV},\;\;M_2=M_3=1\;{\rm TeV},\;\;
M_{(\tilde Q_3, \tilde D_3, \tilde L_3, \tilde E_3)}=1\;{\rm TeV},\;\;
|\mu|=1\;{\rm TeV},\;\;
|A_f|= 1.5\;{\rm TeV}$. Left figure has 
$M_{\tilde U_3}=1$ TeV while right one has
$M_{\tilde U_3}=250$ GeV (the latter giving a rather light stop, ${m}_{\tilde t_1} = 200$ GeV).}
\label{fig:tb20MH1BR}
\end{figure}

To illustrate the validity of this argument, we have selected some specific points from the 
parameter scan to study the deviations in the BR. While doing this we also
wanted to understand the contribution of the different components inducing the 
CP-violating effect. We specifically looked for points with large trilinear 
coupling and large $\mu$ values, for the CP-mixing of the Higgs states 
is proportional to their product (as mentioned in the beginning of this
letter). We then considered different cases choosing the soft-mass values
such that only one of the sparticles in the loop is light, while all others
are heavy (the
mass of the charged Higges boson is varied between 100 GeV and 400 GeV, but 
for the interesting region $M_{H_1}>115$ GeV it is heavier than 300 GeV), 
expecting to see the effect 
due only to the exchange of this light sparticle, alongside the one due to standard
matter, $t$, $b$ and $W^\pm$.
We also considered the situation when only SM particles are effective, with 
all the relevant sparticles heavy.
We plot the BR against $M_{H_1}$ for $\phi_\mu=0^\circ,\;
90^\circ,\;180^\circ$ for two cases, $(i)$ with all sparticles heavy and 
$(ii)$ with a light stop of around 200 GeV, in Fig. \ref{fig:tb20MH1BR}. 
In case $(i)$ the effect is almost entirely due to the CP-mixing of the Higgs 
states, entering the BR through the deviation in the couplings of the SM 
particles with the $H_1$. In case $(ii)$ we also have, in addition to the 
above, the effect of a light stop, through its CP-violating coupling with the 
$H_1$ as well as the sensitivity of the stop mass to the CP-violating phases. 
It is instead found that the effects of light sbottoms, staus and charginos
are negligible, so that the BRs in these cases -- keeping all other SUSY 
parameters to be the same -- are similar to case $(i)$. This is indeed expected, 
considering the smaller Yukawa couplings for the corresponding SM particles 
(with respect to the top) and -- in particular -- the stringent experimental 
limit on $m_{\tilde b_1}$ from Tevatron. The effect in the case of only
heavy sparticles in the loop is exclusively due to the modification of the SM fermion 
and gauge boson couplings of the CP-mixed Higgs state.

\begin{figure}
\includegraphics[width=8cm]{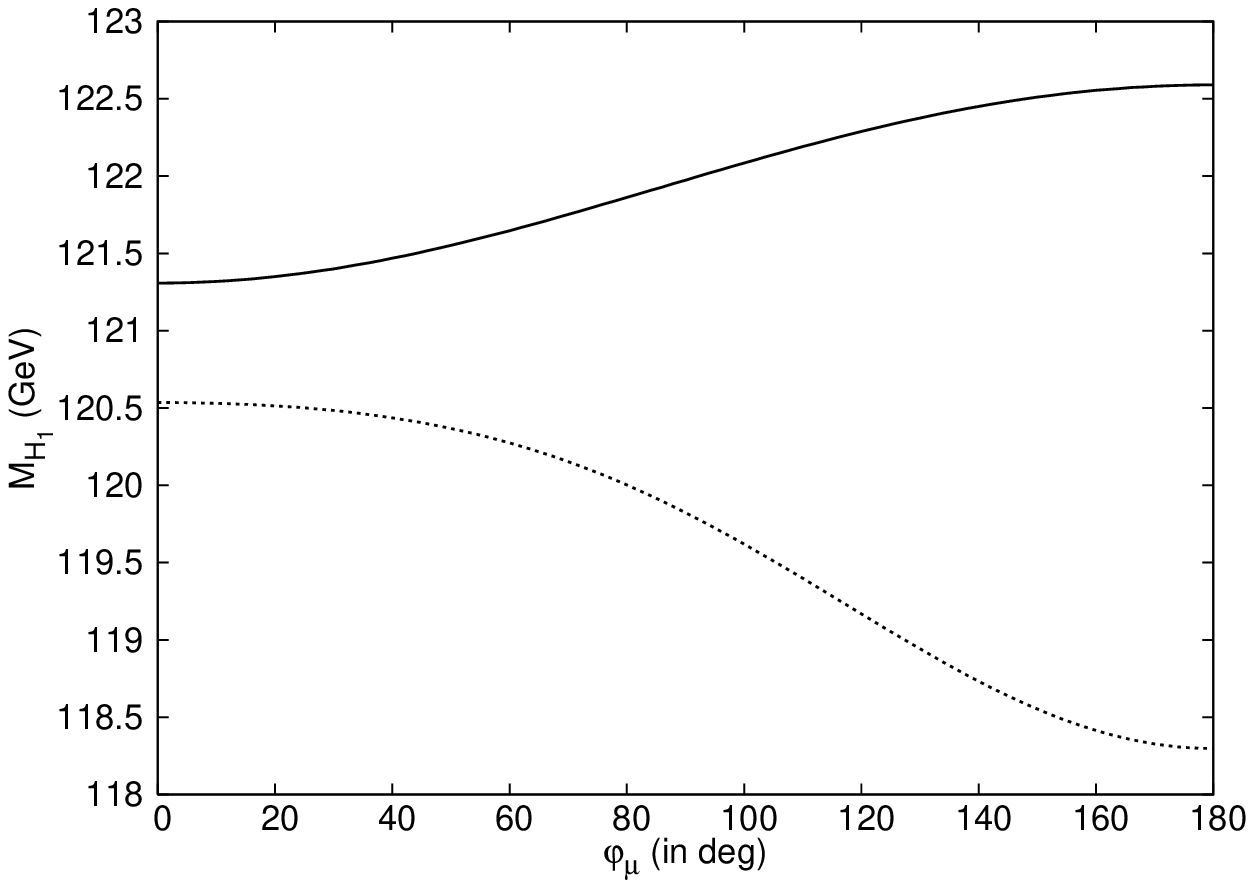}
\includegraphics[width=8cm]{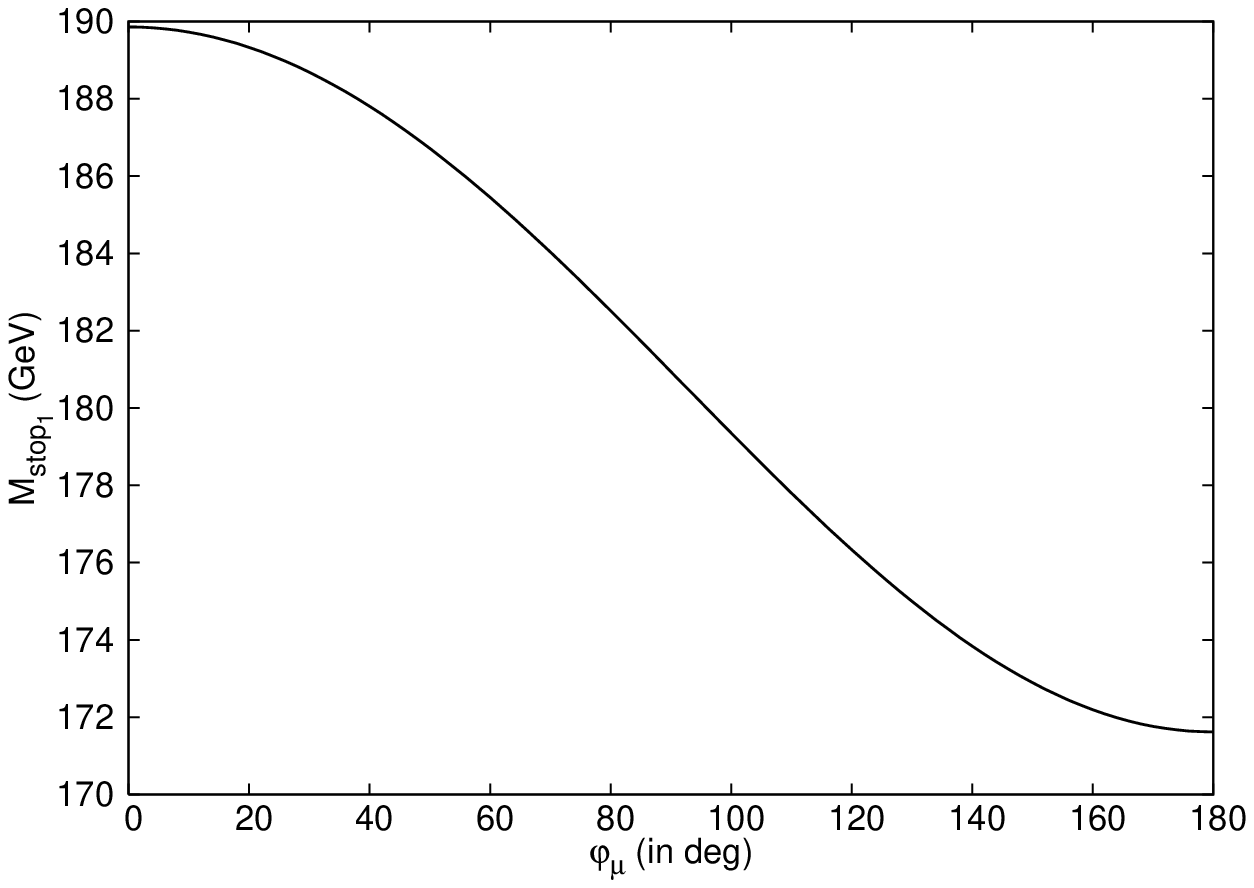}
\caption{$M_{H_1}$ (left) and $m_{\tilde t_1}$ (right) 
plotted against $\phi_\mu$.
In the case of $M_{H_1}$ the solid line corresponds to the case with
$M_{\tilde U_3}=1$ TeV while the dotted line corresponds to 
$M_{\tilde U_3}=250$ GeV. 
Other SUSY parameters are the ones given in Fig.~\ref{fig:tb20MH1BR}.}
\label{fig:tb20MPhi}
\end{figure}

As stated above, the deviations in the $M_{H_1}$ is within 2 GeV for both the 
cases for the entire range of $\phi_\mu$, as shown in 
Fig.~\ref{fig:tb20MPhi} (left). The larger sensitivity of the BR to $\phi_\mu$
in the case with light stop can partly be explained through the sensitivity of 
$m_{\tilde t_1}$ to $\phi_\mu$. The variation of $m_{\tilde t_1}$ with 
$\phi_\mu$ is plotted in Fig.~\ref{fig:tb20MPhi} (right). Notice that
the latter is of the same order as the expected experimental resolution 
\cite{ATLASTDR2,CMS},
so that it may not be possible to confirm CP-violating effects directly
in the stop sector.

In summary, while a full study incorporating the production processes and detector 
dependent aspects is needed to have a clear quantitative picture, our preliminary analyses indicate that the 
di-photon channel of the lightest Higgs boson may enable one to distinguish the
CP-violating MSSM from the CP-conserving one, so long that some SUSY parameters 
are measured elsewhere. This is not phenomenologically unconceivable, as the $H_1\to\gamma\gamma$ detection
mode requires a very high luminosity, unlike the discovery of those sparticles (and the measurement of
their masses and couplings) that impinge on the Higgs process studied here. A complete analysis
will eventually require to fold the decay process with propagator effects and the appropriate
production mode (gluon-gluon fusion and Higgs-strahlung in this case), where similar CP-violating
effects may enter.

\section{Acknowledgements}
We thank Stefan Hesselbach for useful discussions. PP's research is supported by the 
Framework Programme 6 via a Marie Curie International Incoming Fellowship, contract number 
MIF1-CT-2004-002989. This research has been partially financed by the NATO Collaborative 
Linkage Grant no. PST.CLG.980066.

\end{document}